\documentclass[aps,prl,twocolumn,a4paper,superscriptaddress]{revtex4}

\usepackage[paperwidth=215mm,paperheight=300mm,centering,hmargin=1.6cm,vmargin=2cm]{geometry}
\usepackage{multibbl}
\usepackage[pdftex]{graphicx}
\usepackage{amsmath,SIunits}
\usepackage[latin1]{inputenc}
\usepackage{latexsym,stmaryrd}
\usepackage[sort&compress]{natbib}
\usepackage[pdftex]{graphicx}
\usepackage{color}

\makeatletter
\newcommand*{\citenst}[2][]{%
  \begingroup
  \let\NAT@mbox=\mbox
  \let\@cite\NAT@citenum
  \let\NAT@space\NAT@spacechar
  \let\NAT@super@kern\relax
  \renewcommand\NAT@open{[}%
  \renewcommand\NAT@close{]}%
  \citep{#2}%
  \endgroup
}
\makeatother

\renewcommand{\figurename}{\textbf{Figure}}

\begin{document}

\title{All-optical radiofrequency modulation of Anderson-localized modes}

\author{G. Arregui}
\affiliation{Catalan Institute of Nanoscience and Nanotechnology (ICN2), CSIC and The Barcelona Institute of Science and Technology, Campus UAB, Bellaterra, 08193 Barcelona, Spain}
\affiliation{Dept. de F\'{i}sica, Universitat Autonoma de Barcelona, 08193 Bellaterra, Spain}
\author{D. Navarro-Urrios}
\affiliation{Catalan Institute of Nanoscience and Nanotechnology (ICN2), CSIC and The Barcelona Institute of Science and Technology, Campus UAB, Bellaterra, 08193 Barcelona, Spain}
\author{N. Kehagias}
\affiliation{Catalan Institute of Nanoscience and Nanotechnology (ICN2), CSIC and The Barcelona Institute of Science and Technology, Campus UAB, Bellaterra, 08193 Barcelona, Spain}
\author{C. M. Sotomayor Torres}
\affiliation{Catalan Institute of Nanoscience and Nanotechnology (ICN2), CSIC and The Barcelona Institute of Science and Technology, Campus UAB, Bellaterra, 08193 Barcelona, Spain}
\affiliation{ICREA - Instituci\'o Catalana de Recerca i Estudis Avan\c{c}ats, 08010 Barcelona, Spain}
\author{P. D. Garc\'{i}a}
\email{david.garcia@icn2.cat}
\affiliation{Catalan Institute of Nanoscience and Nanotechnology (ICN2), CSIC and The Barcelona Institute of Science and Technology, Campus UAB, Bellaterra, 08193 Barcelona, Spain}
\homepage{http://www.icn.cat/~p2n/}

\date{\today}

\small

\begin{abstract}
All-optical modulation of light relies on exploiting intrinsic material nonlinearities~\cite{Silicon}.\ However, this optical control is rather challenging due to the weak dependence of the refractive index and absorption coefficients on the concentration of free carriers in standard semiconductors~\cite{Soref}.\ To overcome this limitation, resonant structures with high spatial and spectral confinement are carefully designed to enhance the stored electromagnetic energy, thereby requiring lower excitation power to achieve significant nonlinear effects~\cite{Notomi_nonlinearities}.\ Small mode-volume and high quality ($\text{Q}$)-factor cavities also offer an efficient coherent control of the light field and the targeted optical properties.\ Here, we report on optical resonances reaching $\text{Q} \sim 10^5$ induced by disorder on novel photonic/phononic crystal waveguides.\ At relatively low excitation powers (below $1\,\text{mW}$), these cavities exhibit nonlinear effects leading to periodic (up to $\sim$ 35 MHz) oscillations of their resonant wavelength.\ Our system represents a testbed to study the interplay between structural complexity and material nonlinearities and their impact on localization phenomena and introduces a novel functionality to the toolset of \textit{disordered photonics}.
\end{abstract}


\maketitle

\begin{figure}[t!]
 \centering
  \includegraphics[width=8.9cm]{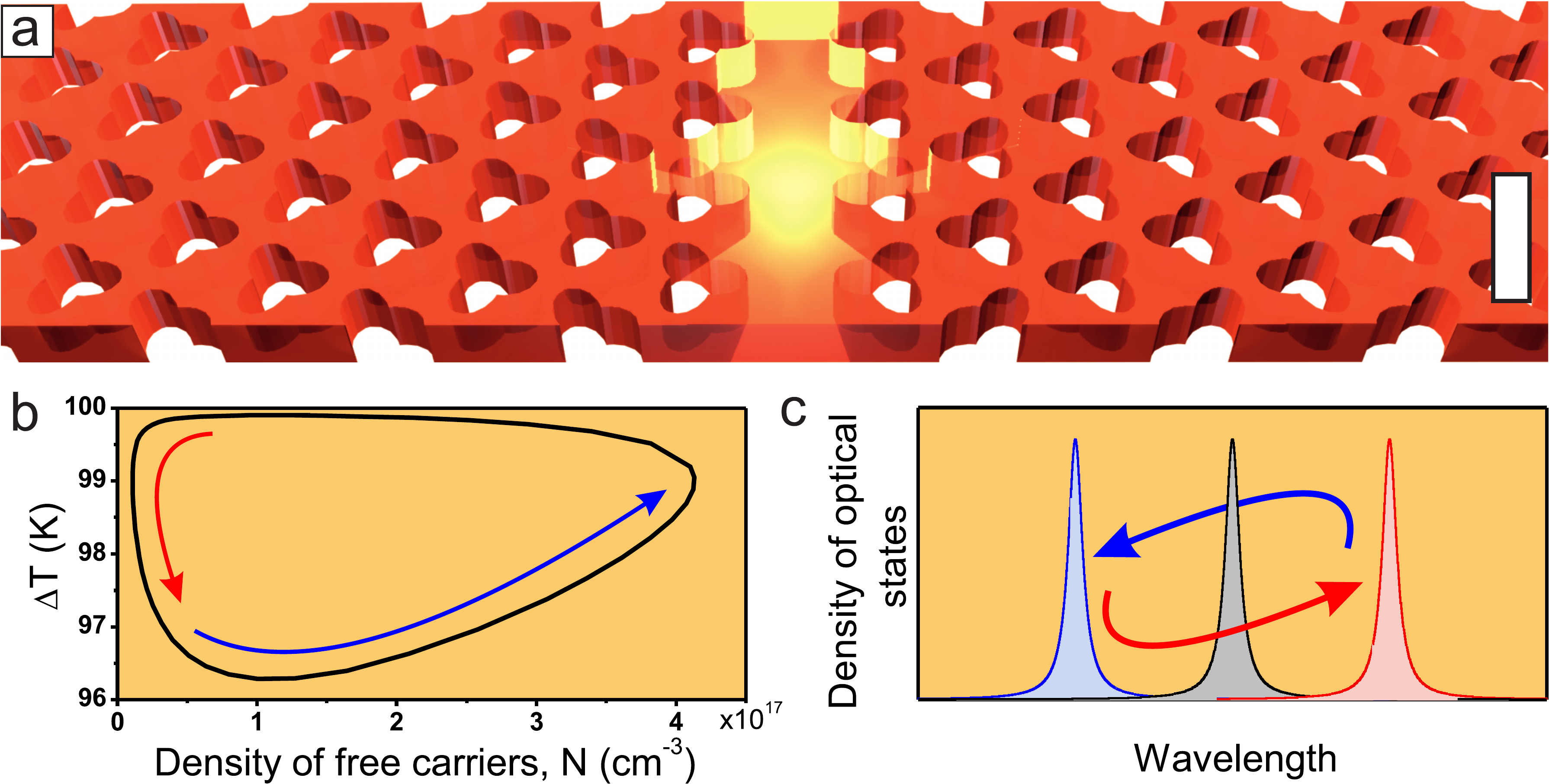}
      \caption{ \label{1} \textbf{Optical nonlinearity in a \emph{shamrock}-crystal waveguide.} (color online) (\textbf{a}) Illustration of a shamrock-crystal waveguide where the building block, the \textit{shamrock}, is formed by overlapping three ellipsoids, rotated by $2 \pi/3$ with respect to each other (details in the supplementary material).\ The residual imperfection due to the fabrication process induces optical confinement along the waveguide.\ The scale bar is $500\,\text{nm}$.\ (\textbf{b}) A limit cycle dynamic solution to the differential equations describing the nonlinear optical system.\ Here, the structural heating,  $\Delta \text{T}$, and the density of free carriers, $\text{N}$, are linked to each other trough the number of photons in the cavity, leading to a stable closed trajectory in phase space.\ (\textbf{c}) The dynamic variables of the system, $\Delta \text{T}$ and $\text{N}$, have an opposite dispersive effect on the refractive index of the structure which induces a temporal modulation of the disorder-induced mode.}
    \end{figure}

A successful strategy to achieve efficient optical confinement consists of introducing controlled point or line defects in otherwise regular dielectric lattices.\ This has applications in both the classical and the quantum phenomena~\cite{review} such as slow light~\cite{Baba}, efficient single-photon sources~\cite{beta-factor}, nanolasing~\cite{Noda_laser}, optomechanical coupling~\cite{Painter} and even enhanced interaction between light and single atoms~\cite{Kimble}.\ A bottleneck of this strategy relies on the extreme sensitivity of particularly high (as designed) $\text{Q}$-factor cavities ($\text{Q} \sim 10^6$) to uncontrolled imperfections appearing during the fabrication process~\cite{Gerace,Savona_optimization}.\ A less typical strategy to confine light consists of exploiting such imperfections~\cite{Annalen}.\ Small spatial fluctuations of the order of few $0.001a$ in the position of the lattice building blocks, where $a$ is the typical lattice parameter, give rise to strong multiple scattering which results in efficient optical confinement by recurrent interference with quality factors~\cite{Vollmer} reaching $\text{Q} \sim 10^6$, thus competing in performance with engineered defects while being inherently robust against disorder.\ Here, we exploit material nonlinearities to modulate these type of disorder-induced optical modes on a novel-design photonic-crystal waveguide which also allows for strong confinement of mechanical motion.\ Fig.~\ref{1}a displays a scheme of a \emph{shamrock}-crystal waveguide, where the building block - the \emph{shamrock} - is obtained by overlapping three ellipsoids rotated by $2 \pi/3$ with respect to each other as detailed in Ref.~\citenst{Immo} and in the supplementary material.\ This structure is similar to a standard photonic-crystal waveguide~\cite{Baba} where the replacement of holes by shamrocks allows accommodating, simultaneously, a waveguide for THz-frequency photons and GHz-frequency phonons for a lattice parameter $a = 500\,\text{nm}$.\ We fabricate our structures in silicon where material nonlinearities result~\cite{thermooptic} in an excess of free carriers and a significant structural heating under high optical excitation.\ As sketched in Fig.~\ref{1}b and c, these two processes become linked to each other while inducing an opposed dispersive effect on the refractive index of the structure, which eventually leads to closed stable trajectories in phase space and the periodic modulation of disorder-induced cavities at high radio frequency.

\begin{figure*}
\centering
 \includegraphics[width=18.4cm]{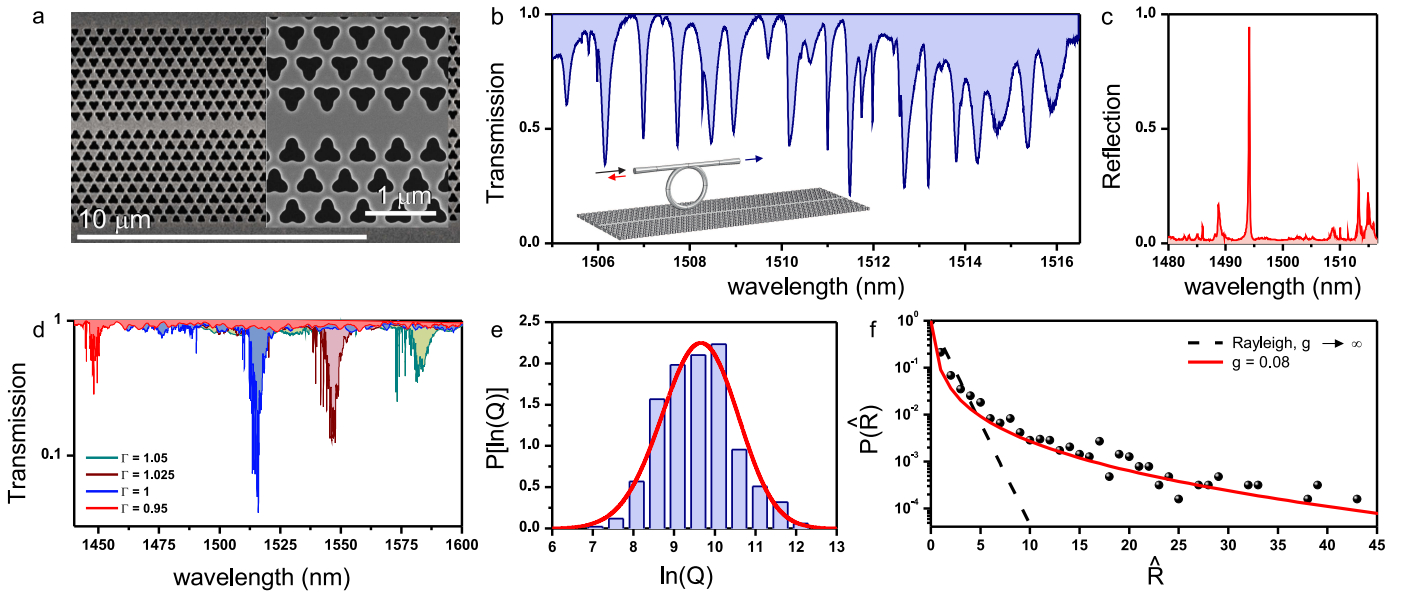}
    \caption{ \label{2} \textbf{Disorder-induced localization in shamrock-crystal waveguides.} The optical modes of the structure are probed through a tapered-fiber loop which is carefully placed in close proximity to the waveguide.\ (\textbf{a}) Scanning electron micrograph of a silicon shamrock-crystal waveguide obtained by etching two shamrock crystals with opposed symmetry and lattice constant $a = 500\,\text{nm}$.\ (\textbf{b}) Light from a tunable diode laser is transmitted trough the fiber and couples evanescently with the localized modes of the structure, revealed as dips in transmission and peaks in reflection (\textbf{c}).\ (\textbf{d}) These localized modes appear in the slow-light regime of the waveguide and are red-shifted when the entire in-plane nanostructure is scaled by a factor $\Gamma$.\ (\textbf{e}) Histogram of the experimental $\text{Q}$-factors observed along the waveguide.\ The red curve plots the calculated distribution with parameters $\xi/\text{L}=0.045\pm0.015$ and $\ell/\text{L}>10^5$, where $\xi$ is the localization length, $\ell$ is the loss length and $\text{L}=150\,\micro\text{m}$ is the total length of the waveguide (more details in supplementary information).\ (\textbf{f}) Reflectance probability distribution where $\hat{\text{R}}$ is the measured reflectance normalized to its average $\langle \text{R}\rangle$.\ The solid line represents the best fit to the theory of Ref.~\citenst{Rossum} with the dimensionless conductance, $\textit{g}$, as the single fitting parameter.}
\end{figure*}

The optical modes induced by disorder in a shamrock-crystal waveguide, as the one shown in Fig.~\ref{2}a, appear around the slow-light region of the dispersion relation~\cite{Vollmer}, in our case $\sim 1500\,\text{nm}$ for $a = 500\,\text{nm}$.\ These modes are revealed in the transmitted light trough a tapered-fiber loop~\cite{Painter} when tuning a diode laser across this spectral region, as plotted in Fig.~\ref{2}b.\ When the fiber is placed in close proximity to the waveguide, as sketched in the inset of Fig.~\ref{2}b, the dips in transmission correspond to disorder-induced or Anderson-localized optical modes along the waveguide evanescently coupled to the fiber loop.\ These modes appear as peaks in the reflected signal measured with a $50/50$ fiber beam-splitter in the fiber input, as shown in Fig.~\ref{2}c.\ The band~\cite{Lifshitz} of Anderson-localized modes can be frequency-tuned by scaling the full in-plane structure by a factor $\Gamma$, as shown in Fig.~\ref{2}c.\ These modes present a rather broad distribution of $\text{Q}$-factors - plotted in Fig.~\ref{2}e - with values in the range $1\,\times10^3 < \text{Q} < 1.5\,\times10^5$ and a mean value of $\text{Q}\sim1.6\,\times10^4$.\ By comparing these values with the ones calculated with a fully three-dimensional Bloch mode expansion technique~\cite{Vasco}, we can estimate the amount of fabrication imperfection as $\sigma = 0.006a$, which is very similar to previous estimations of the tolerance of the fabrication process~\cite{quantifying}.\ Here, we map all the possible different sources of intrinsic disorder to zero-mean Gaussian random displacements in the \emph{shamrock} positions with standard deviation $\sigma$.

By analyzing the $\text{Q}$-factor distribution plotted in Fig.~\ref{2}e, we can extract very relevant information of the Anderson-localized modes in our structure.\ The intrinsic values of this log-normal distribution are mainly determined by in-plane finite-size effects along the waveguide~\cite{Smolka,Savona_waveguide}, i.e., by the average extension of the localized modes - $\xi$ - compared to the total length of the system, $\text{L}=150\,\micro \text{m}$.\ Further loss mechanisms such as material absorption or out-of-plane leakage, which we quantify with the waveguide \textit{loss} length $\ell$, may reduce these values limiting the maximum $\text{Q}$-factors achievable.\ In this picture, the ratio $\xi/\text{L}$ determines the log-normal values, while the ratio $\ell/\text{L}$ may impose a truncation to the distribution (see details in the supplementary information).\ By fitting the measured distributions with a truncated log-normal, we obtain $\xi/\text{L}=0.045\pm0.015$ and $\ell/\text{L}>10^4$.\ These values open the possibility for light-matter strong coupling induced by disorder, as calculated in Ref.~\citenst{Tyrrestrup}.\ The spectral fluctuations plotted in Fig.~\ref{2}b and c can be further used to obtain the dimensionless conductance, $\textit{g}$, which quantifies the confinement induced by disorder in our structures.\ The smaller its value, the stronger the confinement.\ Fig.~\ref{2}e plots the probability distribution of the reflectance spectra collected by placing the fiber loop at different positions along the waveguide.\ We do this at the lowest input power ($5\,\micro\text{W}$) to rule out any effect of material nonlinearities.\ The tail of the distribution reveals the presence of few but very bright peaks when compared to the background which is the fingerprint for Anderson localization.\ Fitting this distribution with the theory developed in Ref.~\citenst{Rossum} yields a dimensionless conductance of $\textit{g}=0.08\pm0.01$ as a single fitting parameter, which is lower than in previous experiments~\cite{Tigelen,Annalen} confirming how efficient is the confinement induced by fabrication imperfections.

\begin{figure*}
\centering
  \includegraphics[width=18.4cm]{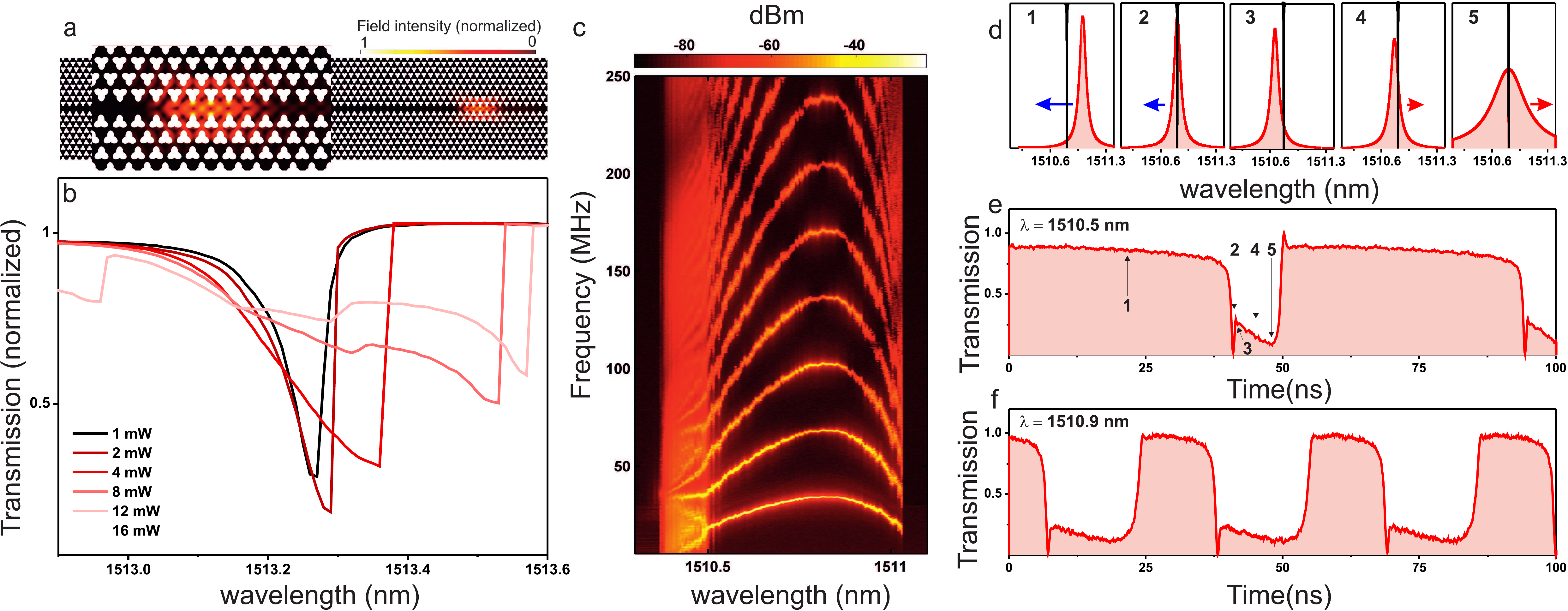}
    \caption{ \label{3} \textbf{Nonlinear modulation of Anderson-localized modes.} (Color online) (\textbf{a}) Light field intensity of a typical eigenmode of a shamrock-crystal waveguide with a lattice constant $a=500\,\text{nm}$ (other parameters detailed in the supplementary material), calculated around the cutoff frequency of the ideal slow-light waveguide mode.\ The position of the shamrocks is randomized normally with a standard deviation of $\sigma=0.006a$ mimicking the effect due to the fabrication process.\ (\textbf{b}) Normalized transmission spectra of an Anderson-localized optical mode measured upon increasing the input power, which shows a typical \emph{sawtooth}-shaped transmission bistability, caused by the material nonlinearities.\ (\textbf{c}) Color contour plot of the radiofrequency power spectral density at an input power of $16\,\text{mW}$ measured while scanning the input laser wavelength, $\lambda_\text{L}$, through an Anderson-localized mode with a \textit{cold} wavelength at $\lambda_\text{c}=1510.3\,\text{nm}$.\ (\textbf{d}) Sketch of the fast modulation cycle of an Anderson-localized mode around $\lambda_\text{L}$ marked with a solid-blue line.\ (\textbf{e}) and (\textbf{f}) time-resolved transmission measured with an input laser wavelength at $\lambda_\text{L} = 1510.5\,\text{nm}$ and $\lambda_\text{L} = 1510.9\,\text{nm}$, respectively.}
\end{figure*}

Large electromagnetic energy stored in a small volume results in a highly nonlinear behavior~\cite{Vahala}.\ The resonant recirculation of weak excitations within an Anderson-localized mode is proportional to the input power and to the ratio $\text{Q}/\text{V}$.\ To estimate their typical mode profile and mode volume, we calculate the eigenmodes of a shamrock-crystal waveguide with the same structural parameters of the fabricated structures using a commercial finite-element solver.\ The total length of the simulated waveguide is $100a$ and both sides of the structure are terminated with reflectionless absorbers in order to mimic an open system.\ The calculated waveguide is perturbed by a zero-mean Gaussian disorder in the position of the shamrocks with standard deviation $\sigma = 0.006a$ to reproduce the effect of the fabrication process (details in the supplementary material).\ A calculated interference pattern along the waveguide is shown in Fig.~\ref{3}a for a wavelength $\lambda_\text{c}=1512.95\,\text{nm}$, revealing a typical Anderson-localized mode with a volume $\text{V}=1.2(\lambda_\text{c}/ n)^3$.\ Here, $\lambda_\text{c}$ is the cavity wavelength in the linear regime - the \textit{cold} wavelength - and $n$ is the material refractive index.\ These mode profile and mode volume are comparable to the ones of a carefully designed heterostructure-cavity mode~\cite{heterostructure}.\ As the intracavity optical-field intensity is enhanced in proportion to $\text{Q}/\text{V}$, a very low input power triggers nonlinear effects in the shamrock-crystal waveguide.\ We confirm this by increasing the input power of the excitation laser in our experiment.\ The transmission spectrum trough the fiber-loop evanescently coupled to an Anderson-localized mode shows a typical \emph{sawtooth}-shaped transmission bistability at an input power typically around $1\,\text{mW}$, as plotted in Fig.~\ref{3}b.\ Sweeping the excitation laser wavelength at higher powers drags the resonant wavelength to higher values, broadening the range of the hysteresis loop.

The main nonlinear process in standard silicon resonators in the telecom spectral range is two-photon absorption which generates free carriers through an indirect-band phonon creation~\cite{thermooptic}.\ This induced free carrier population absorbs additional photons which leads to structural heating when the carriers decay.\ In addition, these two parameters are linked, leading to complex dynamics of the resonant wavelength which can be described by a system of nonlinear coupled rate equations~\cite{rate_equations} (details in supplementary information).\ This phenomenon has been observed in different resonant photonic nanostructures such as microdisks~\cite{rate_equations}, photonic crystals~\cite{nonlinear_phot_crystal} or optomechanical crystals~\cite{dani_selfpulsing} but it is still unexplored in the context of disordered photonics.\ By measuring the oscillatory components of the transmitted light with a fast photodetector while sweeping the laser wavelength, $\lambda_\text{L}$, it is possible to unravel the complex dynamics of a particular Anderson-localized mode with a \textit{cold} cavity wavelength of $\lambda_\text{c}=1510.3\,\text{nm}$ and $\text{Q}=6\,\times10^4$.\ Fig.~\ref{3}c plots this radiofrequency modulation starting with the laser blue-detuned with respect to the Anderson mode.\ For low input laser wavelengths, the resonant wavelength presents a stable temperature-dominated and time-independent red shift.\ Increasing $\lambda_\text{L}$ beyond $1510.5\,\text{nm}$, which corresponds to the intra-cavity photon threshold value~\cite{rate_equations} of $n_\text{0,th} \sim 1.35\,\times10^5$, leads to a periodic modulation of the intra-cavity mode wavelength~\cite{selfpulsing,Chaos}, revealed as a clear frequency comb with increasing spacing.\ To understand the dynamics of this optical modulation, Fig.~\ref{3}d sketches different frames of the period of the Anderson-localized mode oscillation around the laser wavelength marked as a black-solid line.\ Panels~\ref{3}d(1)-(3) plot the first half period of the oscillation when the cavity mode is progressively blue-shifted due to a slow temperature recovery and a fast free-carrier population build-up, as sketched in Fig.~\ref{1}b.\ Far from resonance, at position (1), the transmission trough the fiber at $\lambda_\text{L}$ is maximum due to the poor coupling to the cavity mode.\ When the cavity is tuned on resonance with the laser - position (2) - the transmission trough the fiber drops to the minimum due to the evanescent coupling of the transmitted light to the Anderson-localized cavity.\ The oscillation still continues to the maximum blue-shift at position (3), for which the effect of heating of the sample overcomes the effect of the excess of free carriers and red-shifts the localized mode back to its initial position.

The full oscillation cycle is detected by time-resolving the transmission trough the fiber at a fixed $\lambda_\text{L}$ with a fast oscilloscope.\ As shown in Fig.~\ref{3}e and f, the period decreases from $53.7\,\text{ns}$ at $\lambda_\text{L} = 1510.5\,\text{nm}$ to $30.3\,\text{ns}$ at $\lambda_\text{L} = 1510.9\,\text{nm}$.\ The fastest modulation happens at $\lambda_\text{L} = 1510.9\,\text{nm}$, after which the period decreases until the oscillation is completely lost and the resonant wavelength of the mode returns to its initial value at $\lambda_\text{c} = 1510.3\,\text{nm}$.\ The complex dynamics shown here are due to an intricate interdependence of all the different physical parameters involved in the nonlinear rate equations which are developed in detail in the supplementary material.\ Beyond the temporal modulation of the cavity, the material nonlinearities have an effect in the Anderson localization process.\ The optical nonlinearity has two main effects on the $\text{Q}$-factor of the Anderson-localized modes.\ First, the material two-photon absorption induces a dissipation mechanism which increases the optical leakage reducing the $\text{Q}$-factor of the modes.\ This has been observed in pervious experiments with resonant nonlinearity in standard silicon-photonic structures~\cite{thermooptic,dani_selfpulsing}.\ However, the nonlinear process induces an additional dephasing mechanism in the interference process which leads to Anderson localization~\cite{Kivshar}.\ This decoherence mechanism further reduces the $\text{Q}$-factor and it is expected to fully destroy the localization effect in an infinite system~\cite{Cherroret}.\ As sketched in Fig.~\ref{3}d, the $\text{Q}$-factor of the Anderson-localized mode is reduced from position (2) to position (5).\ This \emph{fast} modulation of the $\text{Q}$-factor explains why the two resonant conditions shown in Fig.~\ref{3}e and f do not drop to the same value, a feature that cannot be explained by solely considering the dissipation induced by two-photon absorption and free-carrier induced absorption.\ The multiple scattering process adds additional complexity to this picture.\ When few modes overlap spectrally and spatially, a complex collective behavior is expected in the dynamics of the system due to their interaction.\ The spectral shift induced by the material nonlinearity is strongly mode-dependent which allows tuning different modes on resonance.\ This spatial and spectral mode overlap should result in complex temporal transmission traces that cannot be explained with a single-cavity model.\ However, we do not observe these interaction effects in our experiment.\ We attribute this to the suppression of mode interaction in the localization regime, even between modes that have significant spatial and frequency overlap, as predicted in Ref.~\citenst{interactions} and confirmed by experiments on multimode lasing in the Anderson localized regime~\cite{Jin}.

In summary, we explore here the rich physics resulting from the complex interaction between multiple scattering and material optical nonlinearities.\ We exploit these nonlinearities to modulate resonant modes induced by disorder in the radiofrequency range.\ This all-optical modulation of disorder-induced modes add an extra functionality to the toolset of light-matter interaction mediated by disorder.\ Disordered photonics~\cite{disordered_photonics} offer an alternative and efficient platform for quantum electrodynamics~\cite{Luca} or lasing~\cite{Jin} which are inherently robust against disorder.\ These functionalities, in combination with the fast optical modulation shown here, allow all-optical data-processing granted by imperfection rather than careful engineering.\ While the fundamental limits of engineered optical nanocavities are well known, the limits of this approach are still to be explored, however the performance can be improved even further by inducing long-range disorder correlations~\cite{correlations}.\ Although not shown here, the shamrock nanostructure offers the possibility to couple the electromagnetic field with confined mechanical modes in the GHz-frequency range and to explore optomechanical effects in complex media.

\section{Appendix: Materials and Methods}

\subsection{Sample design and fabrication}

The shamrock-crystal waveguides were fabricated with standard silicon-on-insulator wafers with a top silicon layer thickness of $250\,\text{nm}$ and a $2\,\text{\micro m}$-thick buried oxide sacrificial layer.\ The fabrication of the structure is based on an electron beam, direct-writing process performed on a coated $100\,\text{nm}$ poly-methyl-methacrylate resist film.\ The final shamrock-crystal waveguides are obtained by reactive-ion etching and isotropic vapor etching using hydrofluoride acid solutions.\ The total refractive index of the structure is $\text{n}=3.4$.\ A batch of samples were fabricated with a lattice constant $\textit{a}=500\,\text{nm}$, and a range of scaling factors $\Gamma= [0.95, 1.00, 1.025, 1.05]$ and Gaussian disorder in the position of the shamrocks with standard deviations $\sigma= [0, 0.1a, 0.2a, 0.3a]$.\

\subsection{Experimental setup and optical characterization}

A tapered-fiber loop with a diameter of $\sim 30\,\micro\text{m}$ is placed nearly parallel to the shamrock-crystal waveguide to excite its localized TE-like optical modes.\ A diode laser tunable within $1400\,\text{nm} < \lambda < 1600\,\text{nm}$ and a spectral resolution of $1\,\text{pm}$ is coupled to the fiber input and the transmitted light is detected at the output using two strategies.\ To measure the transmission spectrum, the light is sent directly to a slow photodetector while the reflected light is measured by coupling the photodetector to a $50/50$ fiber beam-splitter in the fiber input.\ The radiofrequency modulation of the transmitted light is measured with an InGaAs fast photodetector with a bandwidth of $12\,\text{GHz}$.\ The radiofrequency voltage is connected to the $50\,\text{Ohm}$ input impedance of a signal analyzer with a bandwidth of $13.5\,\text{GHz}$.\ The transmitted light is time-resolved with an $4\,\text{GHz}$ oscilloscope.\ The whole setup operates at atmospheric conditions of temperature and pressure (see extended detailed information in the supplementary material).\

We measure of a total of 25 input-output reflectance and transmittance curves by varying the position of the tapered-fiber loop along the shamrock-crystal waveguide at the lowest input power of $0.2\,\micro\text{W}$.\ The loaded quality factor $\text{Q}_\text{load}$ is extracted by fitting the transmission dips with a Lorentzian function.\ From this experimental $\text{Q}_\text{load}$, we extract the intrinsic $\text{Q}$-factor as $\text{Q}_\text{load}=2\text{Q}/(1 \pm \sqrt{\text{T}_\text{min}})$, where $\text{T}_\text{min}$ refers to the normalized transmission on resonance with the cavity~\cite{Q_factor_fiber}.\ The full $\text{Q}$-distribution data set is plotted in Fig.~\ref{2}d.

The reflectance probability distribution $\text{P}(\widehat{\text{R}})$ plotted in Fig.~\ref{2}e is measured by collecting the reflectance intensity, $\text{R}_{\text{x},\lambda}$, at different positions of the waveguide, x, while scanning the laser wavelength, $\lambda$.\ Finally, $\widehat{\text{R}}$ is obtained at each position by subtracting the background and normalizing the resulting spectrum by the value averaged over the spectral range $1505\,\text{nm} - 1520\,\text{nm}$.\ Ergodicity on position and wavelength is assumed to obtain the final histogram plotted in Fig.~\ref{2}e.

\section{Acknowledgements}

This work was supported by the Spanish MINECO via the Severo Ochoa Program (Grant SEV-2013-0295) and the project PHENTOM (Fis 2015-70862-P), as well as by  the CERCA Programme / Generalitat de Catalunya, and by the European Commission in the form of the H2020 FET Open project PHENOMEN (GA. Nr. 713450).\ GA is supported by a BIST PhD. Fellowship and  P.~D. Garc\'{i}a and D. Navarro -Urrios gratefully acknowledge the support the Ramon y Cajal fellowship (RYC-2015-18124) and (RYC-2014-15392), respectively.


\begin{widetext}

\newpage

\renewcommand{\figurename}{\textbf{Supplementary Figure}}
\makeatletter
\renewcommand{\thefigure}{S\@arabic\c@figure}
\makeatother
\renewcommand\theequation{S\arabic{equation}}
\renewcommand\thetable{S\arabic{table}}
\renewcommand{\bibname}{References}

\section{supplementary information}

\setcounter{figure}{0}

\section{S1. Shamrock-crystal waveguides: design and fabrication.}

The shamrock crystals~\cite{Immo} used in this manuscript are formed by a triangular lattice of shamrocks with a lattice unit of $a=500\,\text{nm}$.\ A shamrock results from overlapping three ellipsoids rotated by $2 \pi/3$ with respect to one another where the minor and major axis are $0.45a$ and $0.6a$, respectively.\ The center of each ellipsoid is shifted outwards along its major axis by $0.17a$, resulting in a final shape that is reminiscent of a \emph{shamrock}.\ The symmetries and the irreducible Brillouin zone of this crystal are discussed in detail in Ref.\citenst{Immo}.\ By creating this lattice, a TE-like bandgap is open in the THz-frequency range for photons and a full bandgap for GHz-frequency phonons.\ The shamrock-crystal waveguide is engineered by placing two shamrock crystals with opposed symmetry, i.e., with the shamrocks facing each other, separated by $0.9a\sqrt{3}$ distance.\ This distance is crucial to flatter the dispersion relation, i.e., to maximize the group index at the cutoff frequency.

Disorder is introduced by randomizing the position of the shamrocks by an amount, $\Delta \text{\textbf{r}}$, normally distributed with a standard deviation $\sigma = \sqrt{\langle \Delta \text{\textbf{r}}^2\rangle}$ and $\langle \Delta \text{\textbf{r}} \rangle = 0$, where the brackets indicate the expectation value.\ Five sets of waveguides with different scaling factors from $\Gamma = 0.95$ to $\Gamma = 1.05$ were fabricated with disorder ranging from $\sigma=0$ to $\sigma=0.03a$ in steps of $\sigma=0.01a$.\ A total of 20 waveguides were measured.\ Here, $\sigma=0$ implies a perfect structure although the fabrication process always leads to a residual amount of imperfection which is a combination of different types of deviations~\cite{minkovthesis} in the aspect ratio due to overething/underetching, edge/surface roughness, position and shape of the shamrocks.\ To quantity the disorder solely due to the fabrication process, we map all these different types of imperfection as normal random displacements of perfect shamrocks with a standard deviation $\sigma$.\

\begin{figure}[h!]
  \includegraphics[width=14cm]{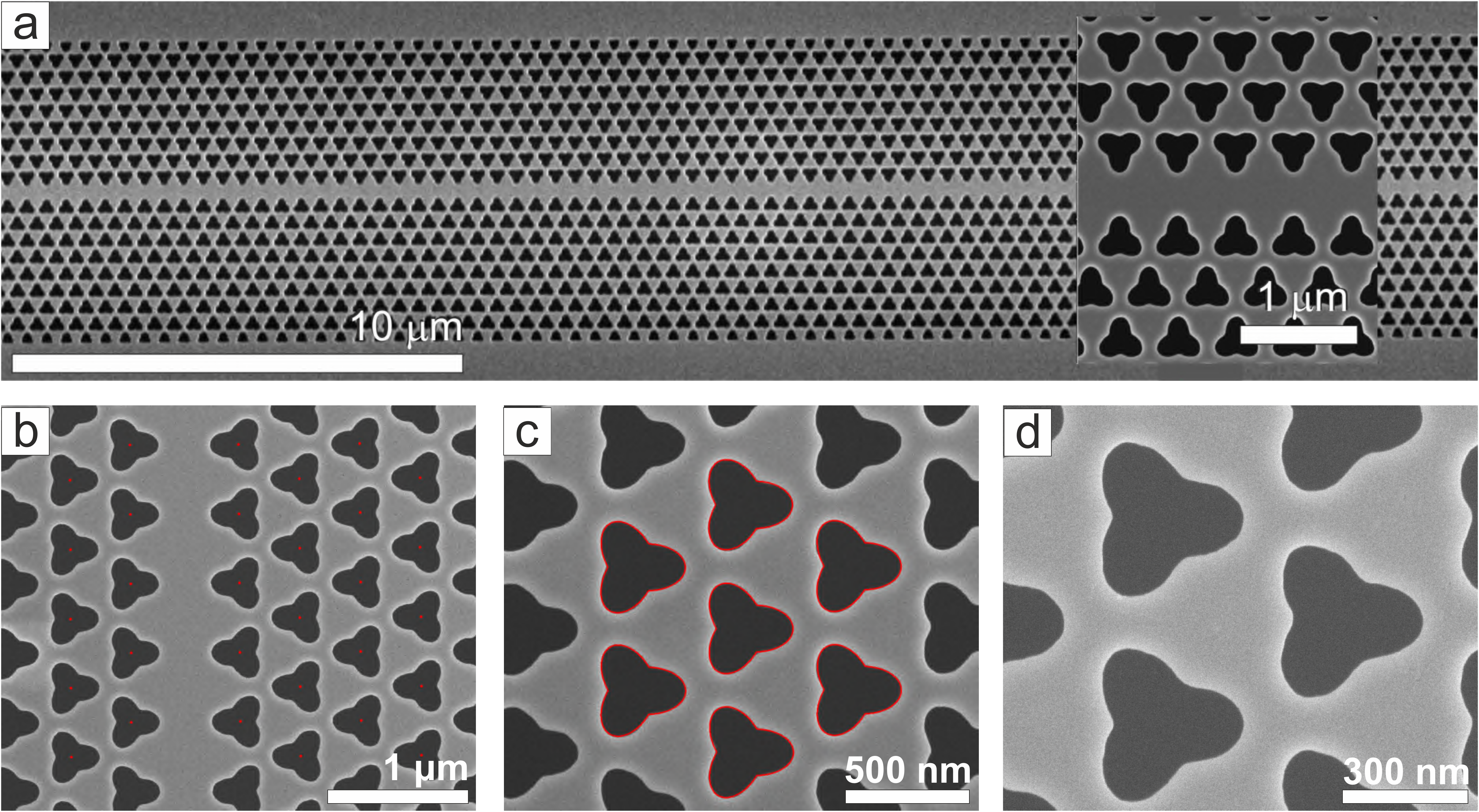}
    \caption{ \label{S1} Scanning-electron micrographs of the shamrock-crystal structure at different scalings.\ Image fits at different levels are included in red.}
\end{figure}

The samples were fabricated in silicon-on-insulator (SOI) wafers with $250\,\text{nm}$-thick silicon, $2\,\text{\micro m}$-thick buried oxide sacrificial layer and $2\,\text{mm}$-thick silicon substrate.\ The wafer was first cleaned in a Piranha solution and dehydrated.\ A Hexamethyldisilazane layer was spin-coated as an adhesion promoter previous to spin-coating a $100\,\text{nm}$-thick poly-methyl-methacrylate (PMMA) resist film.\ These layers were baked at $180\,^{\circ}{\rm C}$ (6 min) for electron-beam exposure.\ The PMMA layer was exposed to an electron beam (Vistec EBPG5000plusES).\ The exposed areas were developed in an IPA:H$_{2}$O (7:3) solution and the PMMA pattern was then transferred to the SOI chip using a reactive-ion etching (Alcatel AMS-110) following the Bosch process with SF$_{6}$ and C$_{4}$F$_{8}$ gases with flows rates of 150 and $100\,\text{sccm}$, respectively.\ Isotropic etching in a saturated hydrofluoride (HF) acid vapor ambient was carried out with a 50\% HF solution during 6 minutes to remove the sacrificial layer and release the $250\,\text{nm}$-thick patterned silicon membranes.\ Figure~\ref{S1} displays scanning electron micrographs (SEM) of the fabricated structures.\ These micrographs were used to extract the mean values of the structural parameters for the bandstructure calculation with a standard image-processing technique and a least-mean squares fitting approach to the shape of the shamrocks.\ Different parameters such as the shape of the shamrocks or the lattice periodicity were extracted from micrographs at different magnifications, as shown in Fig.~\ref{S1}.\ Fig.~\ref{S1}(c) details the roughness of the sidewalls surface.\ However, the resolution of the SEM micrographs limits the reconstruction of actual dielectric profiles and the intrinsic disorder amount is assessed accurately using optical measurements, as described in the main text.

\section{S2. Bandstructure and mode profile calculations.}

\begin{figure}[h]
  \includegraphics[width=11cm]{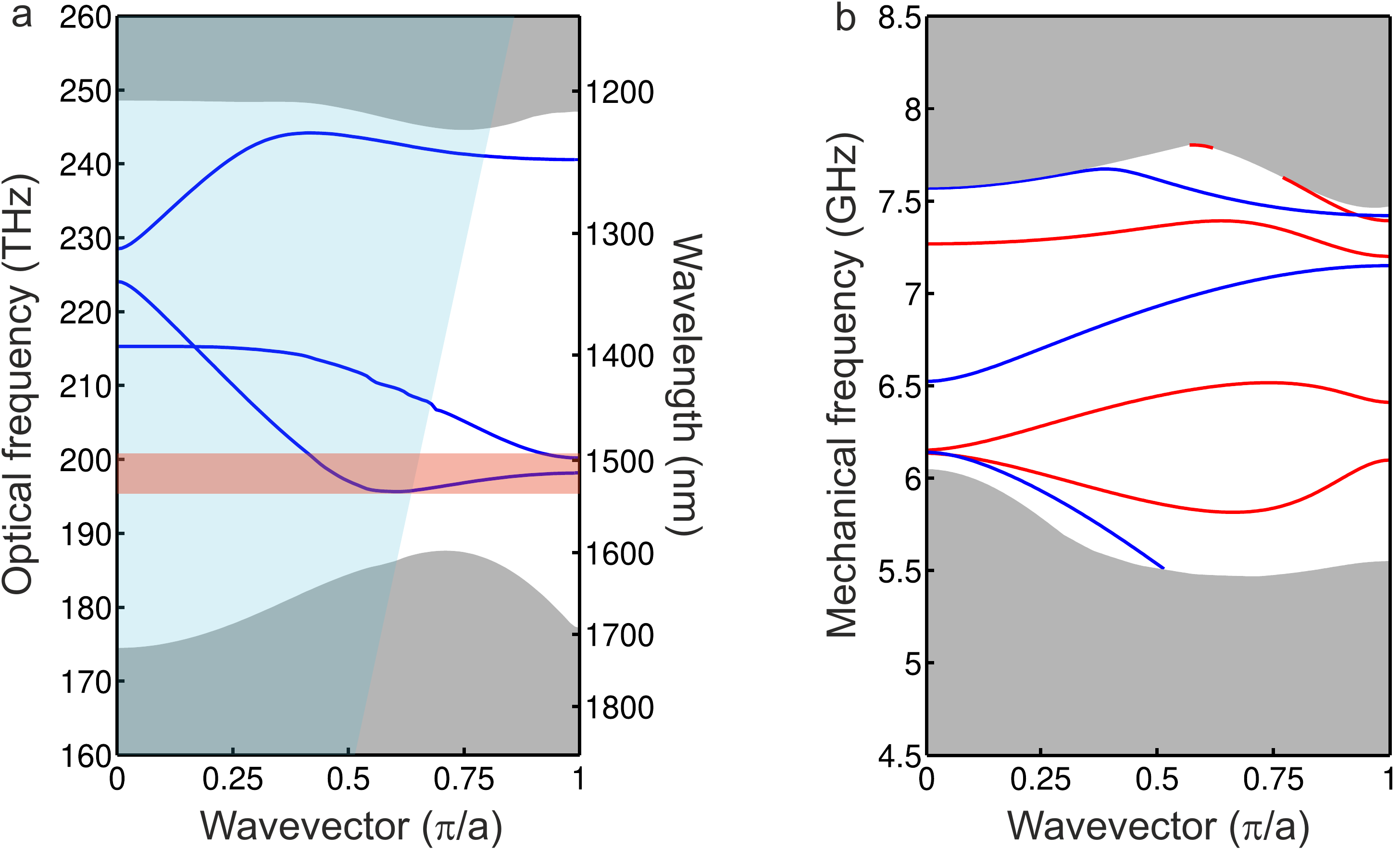}
    \caption{ \label{S2} (Color online) (\textbf{a}) A crystal with lattice constant $a = 500\,\text{nm}$ opens a bandgap for TE-like electromagnetic slab modes around $\sim 200\,\text{THz}$.\ (\textbf{b}) The same crystal opens a full mechanical bandgap around $\sim 7\,\text{GHz}$.\ Several photonic and phononic guided modes are allowed when a controlled line defect is introduced in the structure.\ The blue lines represent z-symmetric guided modes, while the red lines in (\textbf{b}) are z-antisymmetric guided modes.}
\end{figure}

The mechanical and electromagnetic bandstructures of the shamrock-crystal waveguides, plotted in, Figure~\ref{S2}, are calculated using a commercially available finite-element eigensolver - COMSOL Multiphysics - with the mean parameters extracted from the image processing of the fabricated structures.\ The extension and mode volume of the optical and mechanical modes of the structure are calculated with the same eigenmode solver.\ To reproduce the fabricated structures, the waveguides are perturbed by $\sigma=0.006a$ with total length of the simulation domain being $100a$.\ This value of the residual fabrication disorder is obtained by comparing the experimental $\text{Q}$-factor distributions with the calculated using a full three-dimensional Bloch mode expansion technique~\cite{Vasco}.\ To reduce the computational demand~\cite{effective_refractive_index}, it is possible to effectively calculate the $250\,\text{nm}$-thick photonic-crystal slab waveguide bandstructure around the cutoff frequency by simulating a two-dimensional structure with an effective refractive index of $n=2.85$.\ Perfect reflectionless absorbers over a lattice unit at both waveguide terminations were used in order to mimic an open system.\ An eigenfrequency analysis was performed to obtain the photonic and mechanical eigenmodes of the system around the cutoff frequency of the perfect structure, their electromagnetic-field intensity and the total mechanical displacement, respectively.\ Figure~\ref{S3}a plots the electromagnetic-field intensity for different Anderson-localized modes with increasing frequency .\ It shows how their average length increases when its frequency approaches the cutoff frequency of the ideally guided mode, as in standard photonic crystal waveguides~\cite{two_regimes}.\ An heterostructure cavity is simulated based on the design proposed in Ref.~\citenst{heterostructure} for comparison.\ The electromagnetic-field intensity of a photonic eigenmode and the total mechanical displacement field of a phononic eigenmode are plotted in Figure~\ref{S3}(c).\ Remarkably, the smallest mode-volumes induced by disorder are comparable to the mode volume of this ultrahigh-Q engineered cavity.

\begin{figure}[b!]
  \includegraphics[width=16cm]{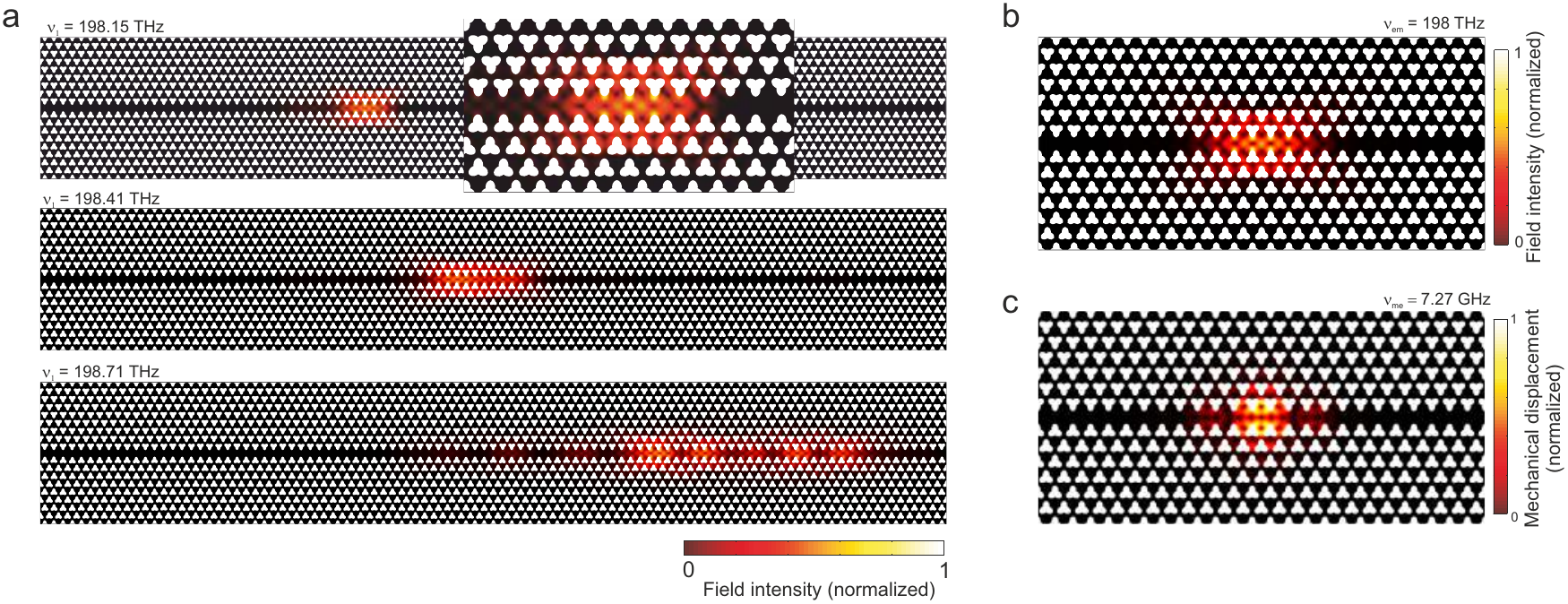}
    \caption{ \label{S3} (Color online) (a) Electromagnetic-field intensity calculated for different Anderson-localized modes with increasing frequency in a shamrock-crystal waveguide perturbed by positional disorder with $\sigma = 0.006a$.\ (b) and (c) plot the electromagnetic-field intensity and the total mechanical displacement, respectively, of a photonic and a phononic eigenmode of an heterostructure cavity calculated as in Ref.~\citenst{heterostructure}.}
\end{figure}

\section{S3. Universality of the Q-factor distribution.}

In random media, the photon decay rate, $\kappa  = \omega /\text{Q}$, shows a universal scaling behavior~\cite{Q_1}, where $\omega$ is the resonant frequency of the mode and $\text{Q}$ is the confinement or quality factor.\ Universality is a remarkably strong property~\cite{Sanli_phd} which implies that the distribution of $\kappa$ or any derived quantity only depends on a single transport parameter, e.g., $\xi/\text{L}$ or any other parameter that characterizes the transport regime of the system.\ Here, we select $\xi/\text{L}$ since it also offers relevant information of the average extension of the cavities induced by disorder.\ The physical picture behind universality implies that two different random systems with different macroscopic and microscopic structure will show the same distribution of $\text{Q}$-factors when both are characterize by the same value for $\xi/L$ as well.

\begin{figure}[t!]
  \includegraphics[width=12cm]{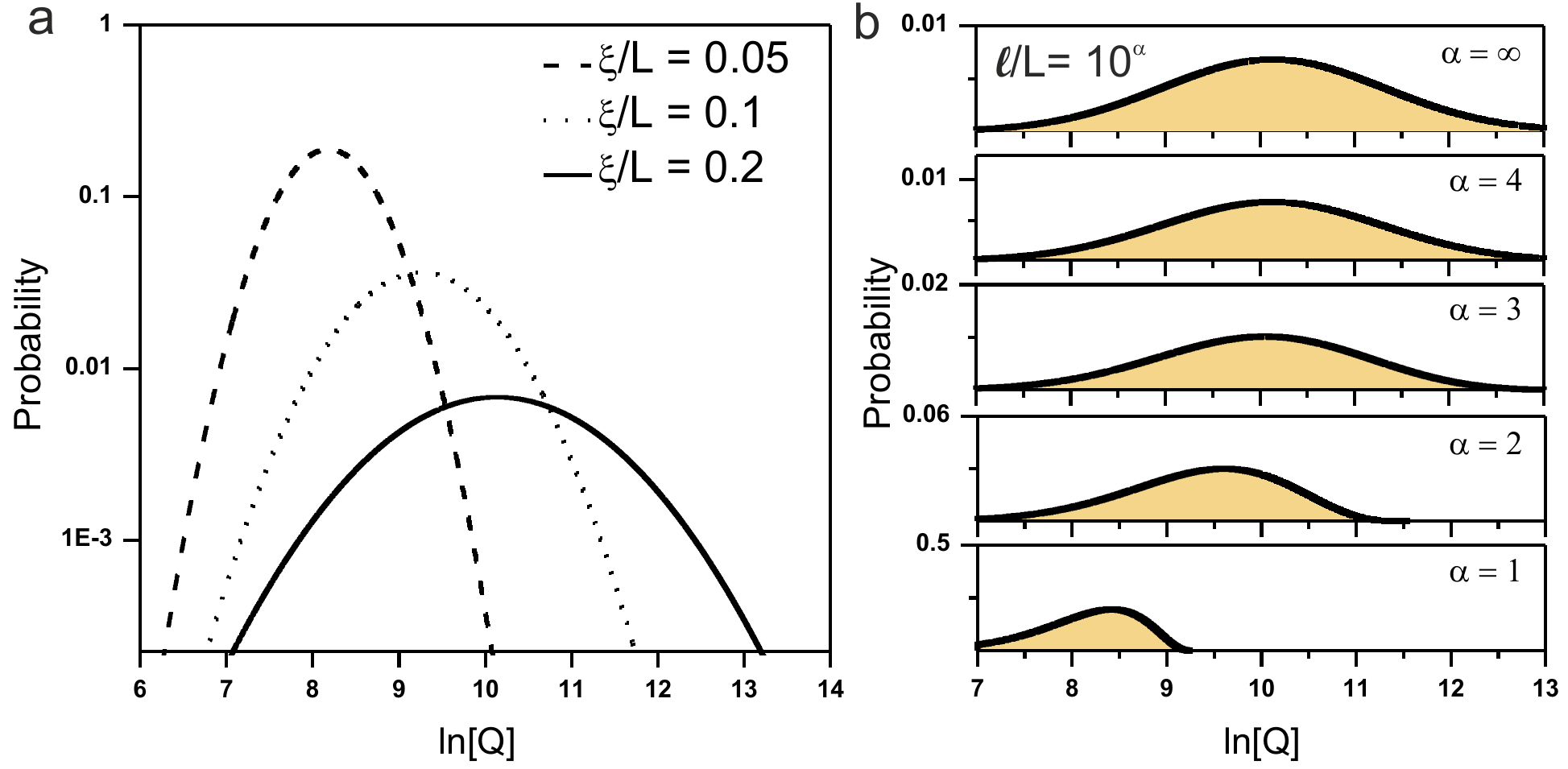}
    \caption{ \label{S4} (Color online) (a) Normalized $\text{Q}$-factor distribution calculated for different localization length ($\xi/\text{L}$) and in absence of out-of-plane loss, i.e., $\ell/\text{L} \rightarrow \infty$.\ (b) $\text{Q}$-factor distribution calculated for a fixed localization length, $\xi/\text{L} = 0.005$, and different values of the loss length expressed in terms of the parameter $\alpha$ as $\ell/\text{L} = 10^\alpha$.}
\end{figure}

In the limit of narrow resonances, the $\text{Q}$-factor is log-normal distributed in the localized regime~\cite{Q_1,Q_2}.\ This relies on the assumption that there are no long-range correlations and the modes are exponentially localized with a gaussian distributed decay length.\ Figure~\ref{S4}a plots the log-normal distributions for the $\text{Q}$-factor as:
\begin{equation}\label{q_factor_distribution}
\text{P}(\text{Q}_0)= \frac{1}{\sqrt{2 \pi} \text{Q}_0 \sigma} e^{-\frac{(\mu - \text{ln}(\text{Q}_0))^2}{2 \sigma^2}}.
\end{equation}
where $\text{Q}_0$ is the in-plane or intrinsic $\text{Q}$-factor and $\mu$ and $\sigma$ are the distribution parameters which only depend on $\xi/\text{L}$.\ Using this rather powerful property of the $\text{Q}$-factor distribution, we can extract the value of $\xi/\text{L}$ from the measured $\text{Q}$-distributions.\ In our structures, however, $\xi/\text{L}$ is not enough to fully describe the experimental $\text{Q}$-factor distributions.\ Other leaky channels such as out-of-plane scattering or material absorption should be taken into account in addition to the in-plane losses~\cite{Savona_waveguide}.\ These extra loss mechanisms may further reduce the value of the $\text{Q}$-factors in the system.\ This out-of-plane leakage is incorporated to our model by adding a generic loss length, $\ell/\text{L}$, which represents a truncation to the log-normal distribution expressed in Eq.~\ref{q_factor_distribution}.\ A detailed analysis of this procedure can be found in Ref.~\citenst{Smolka} and a thorough derivation of the truncated log-normal distribution derivation can be found in section 3.3.1 in Ref.~\cite{Henri_phd}.\ The final truncated $\text{Q}$-distribution is:

\begin{equation}\label{q_factor_distribution_truncated}
\text{P}(\text{Q})= \frac{1}{\sqrt{2 \pi} \sigma} e^{-\frac{ \left[  \mu - \text{ln} \left( \frac{\text{Q} \text{Q}_\ell}{\text{Q}_\ell - \text{Q}} \right) \right] ^2}{2 \sigma^2}} \frac{\text{Q}_\ell \theta (\text{Q}_\ell - \text{Q})}{\text{Q} (\text{Q}_\ell - \text{Q})}
\end{equation}
where $\text{Q}$ is the effective quality factor including in-plane and out-of-plane losses and relates to the intrinsic in-plane $\text{Q}_0$-factor as~\cite{Smolka} $\text{Q}^{-1} = \text{Q}_0^{-1} + \text{Q}_\ell^{-1}$.\ Here, $\text{Q}_\ell$ represents the limit to the highest value of the $\text{Q}_0$-factor that can be measured in the system, imposing a truncation to its log-normal distribution.\ The effect of this truncation is shown in Fig.~\ref{S4}b for different values of $\ell/\text{L}$ expressed in terms of the parameter $\alpha$ as $\ell/\text{L} = 10^\alpha$, and a fixed value of $\xi/\text{L} = 0.05$.\

\begin{figure}[b!]
  \includegraphics[width=10cm]{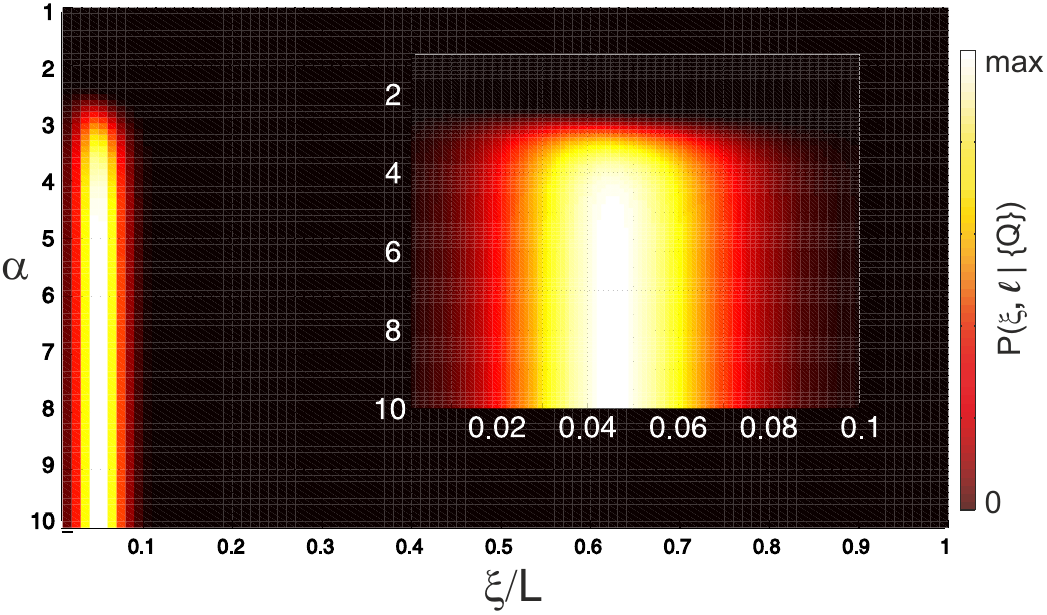}
    \caption{ \label{S5} (Color online) Conditional probability $\text{P}(\text{Q}^m_{i} | \xi,\text{Q}_\ell)$ that the $\text{Q}$-factor distributions measured in the shamrock-crystal waveguide can be described by a localization length $\xi/\text{L}$ and average loss length $\ell/\text{L} = 10^\alpha$.}
\end{figure}

A Bayesian inference~\cite{Bayesian} approach is used to calculate the probability that $\xi/\text{L}$ and $\ell/\text{L}$ determine the set of measured values of the $Q$-factor, $\{ \text{Q}^m_{i} \}$.\ In detail, this is done by calculating the total likelihood:
\begin{equation}\label{Bayesian}
\text{P}(\{ \text{Q}^m_{i}   \} | \xi,\text{Q}_\ell)= \prod^{N}_i \text{P}(\text{Q}^m_{i} | \xi,\text{Q}_\ell)
\end{equation}
where $\text{P}(\text{Q}^m_{i} | \xi,\text{Q}_\ell)$ is given by equation~\ref{q_factor_distribution_truncated} for each individual $\text{Q}$-factor measured in the experiment.\ Figure~\ref{S5} plots the calculated conditional probability that the shamrock-crystal waveguide used in our experiments can be described with a localization length $\xi/\text{L}$ and a loss length $\ell/\text{L} = 10^\alpha$.\ As opposed to standard photonic-crystal waveguides, where only large values of $\text{P}(\text{Q}^m_{i} | \xi,\text{Q}_\ell)$ are observed in a very restricted range~\citep{Smolka}, here the situation is quite different.\ A very large probability is obtained for a very narrow range of $\xi/\text{L}$ and with a lower bound for the loss length ($\ell/\text{L} > 10^4$) which implies a significantly lower out-of-plane scattering (or any other extra leaky mechanisms) in our structures.\ Here, these extra loss channels do not impose a significant truncation to the $\text{Q}$-factor distributions and, therefore, does not represent a significant limitation for the in-plane $\text{Q}$-factors.\ F. Vollmer and co-workers already observed surprisingly narrow cavity resonances reaching $\text{Q} \sim 10^6$ in a pioneering experiment in silicon photonic-crystal waveguides~\cite{Vollmer} and we conclude that the material fabrication and the material itself have a crucial role in the results shown here.\ The range of values of $\xi/\text{L}$ and $\ell/\text{L}$ extracted here are in the regime of light-matter strong coupling, as calculated in Ref.~\citep{Henri_phd}.\ In conclusion, the spectral and spatial confinement induced by fabrication disorder in passive nanophotonic silicon structures are notably competitive with the optical performance of engineered nanocavities.

\section{S4. Analysis of spectral fluctuations.}

Spectral fluctuations in reflection/transmission spectra are a source of information of the transport regime and related parameters in wave propagation.\ Strong fluctuations in the reflection/transmission spectra can only be attributed to the presence of confined modes in the structure and the statistical analysis of these fluctuations is a powerful tool to characterize the regime of localization~\cite{shenglocalization}.\ In particular, the existence of very bright and spectrally well separated peaks in, e.g., the reflectance spectrum is a fingerprint of Anderson localization and its statistical analysis can provide significant information on the degree of confinement induced by disorder in our structures.\ To quantify the degree of confinement, we use a parameter, the dimensionless conductance $\textit{g}$, initially proposed~\cite{g} in the scaling theory as a single scaling parameter to describe the conductor-insulator phase transition induced by disorder.\ This theory was extended to the case of classical waves by Van Rossum and Neuwenhuizen~\cite{Rossum} and defined as the total transmittance, i.e., the sum over all the transmission coefficients connecting all input-output modes.\ It governs all the statistical aspects of light transport in a random medium~\cite{Chabanov} and sets the boundary for Anderson localization in absence of absorption at $(\textit{g} \leq 1)$ for three-dimensional structures~\citep{Tigelen}.\ For lower-dimensional structures, $\textit{g}$ determines the degree of confinement and can be extracted as a single fitting parameter by fitting the transmitted intensity distribution with the theory developed in Ref.~\citep{Rossum}
\begin{equation}\label{intensity_distributoion}
\text{P}(\widehat{\text{I}})= \int_{- i \infty}^{i \infty} \! \frac{\mathrm{d} x}{\pi i} \text{K}_0 (2 \sqrt{-\widehat{\text{I}}x}) e^{-\Phi_{\text{con}}(x)},
\end{equation}
where $\widehat{\text{I}}$ is the normalized intensity by its average value $\langle \text{I}\rangle$.\ This is valid in the regime of perturbative scattering and in the absence of absorption.\ Here, $\text{K}_0$ is a modified Bessel function of the second kind and $\Phi_{\text{con}}(x)$ is obtained by assuming plane-wave incidence to be:
\begin{equation}\label{intensity_distributoion}
\Phi_{\text{con}}(x)= \textit{g} \ln^2 \left( \sqrt{1 + \frac{x}{\textit{g}}} +\sqrt{\frac{x}{\textit{g}}} \right)
\end{equation}

For low values of the parameter $\textit{g}$, Eq.~\ref{intensity_distributoion} leads to a heavy tail of the intensity distribution which can only be explained in terms of Anderson localization.\ In the present experiment, it is more convenient to record the spectra in reflectance at different positions along the waveguide in order to obtain the required statistics due to the absence of interference effects arising from the ring-like modes of the fiber-taper loop.

\section{S5. Nonlinear rate equations.}

Optical micro- and nano-cavities driven in the classical regime are often modelled with a temporal coupled-mode theory, which specific formulation depends on the boundary conditions of the system, i.e., the specific excitation scheme~\cite{CPT}.\ The equation governing the cavity-field dynamics in a photonic-crystal cavity with a resonant wavelength $\omega_{\text{c}}$ excited with an external laser at power $\text{P}_\text{in}$ and at frequency $\omega_{\text{L}}$ trough a bidirectional bus waveguide, in our experiments a tapered-fiber loop, can be written as:

\begin{equation}\label{CPTeq}
\dot{A}= -\left(i\Delta + \frac{\kappa}{2}\right)A -\sqrt{\frac{\kappa_\text{e}}{2} \text{P}_\text{in}}
\end{equation}
where $A$ is the complex amplitude of the temporal envelope of the cavity field, $\Delta = \omega_{\text{c}}-\omega_{\text{L}}$ the spectral detuning, $\kappa$ the overall photonic decay rate and $\kappa_{e}$ the photon decay rate into the extrinsic channel used for excitation.\ The outcoupled field satisfies $A_\text{out}=A_\text{in}+  \sqrt{\frac{\kappa_\text{e}}{2}}A$, where a symmetric bidirectional coupling is considered.\ This equation admits a simple steady state solution, $A_\text{o}$, for a continuous-wave laser drive and an intra-cavity photon number $\text{n}_\text{ph}=\frac{\lvert A_\text{o} \rvert^2}{\hbar\omega_{\text{c}}}$:
\begin{equation}\label{no}
\text{n}_\text{ph} =  \left( \frac{2 \text{P}_\text{in}}{\hbar\omega_{\text{L}}} \frac{\kappa_\text{e}} {\kappa^2} \right) \left( \frac{\kappa^2 } {\Delta^2 + \kappa^2} \right)
\end{equation}\

\begin{figure}[t!]
  \includegraphics[width=12cm]{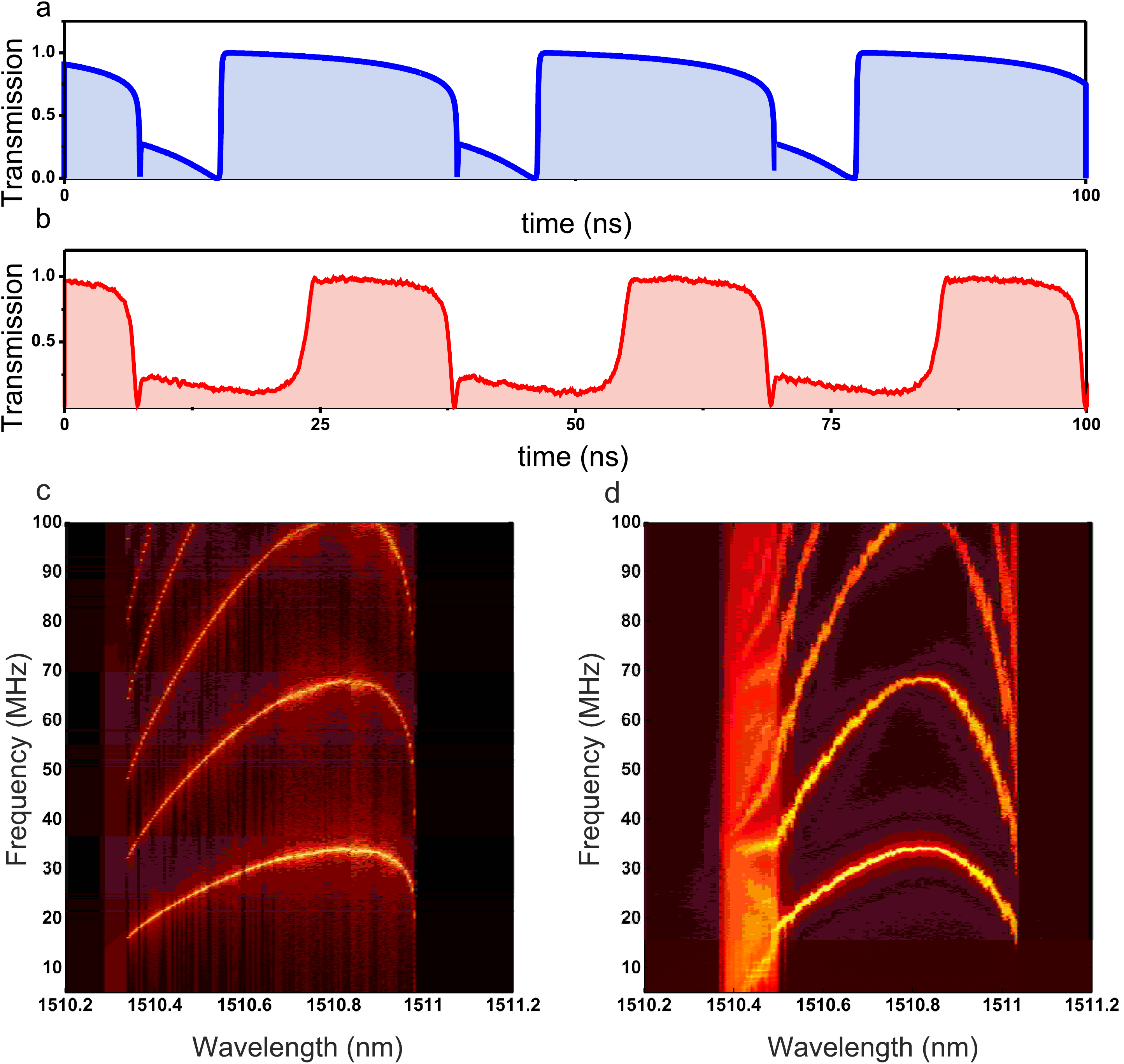}
    \caption{ \label{S6} (Color online) (a) Calculated time-resolved transmission for an optical mode with a \textit{cold} wavelength $\lambda_\text{c}=\,\text{nm}$, an input power $\text{P}_\text{in}=\,\text{mW}$ and an input laser wavelength $\lambda_\text{L}=\,\text{nm}$.\ The rest of the parameters of the simulation are summarized in the table~\ref{table}.\ (b) Measured time-resolved transmission for a Anderson-localized mode with a \textit{cold} wavelength $\lambda_\text{c}=1510.3\,\text{nm}$, an input power $\text{P}_\text{in}=16\,\text{mW}$ and an input laser wavelength $\lambda_\text{L}=1510.9\,\text{nm}$.\ Contour plot of the radiofrequency power spectral density calculated (c) and measured (d) while scanning the input laser wavelength through an Anderson-localized optical mode with a \textit{cold} wavelength $\lambda_\text{c}=1510.3\,\text{nm}$ at an input power $\text{P}_\text{in}=16\,\text{mW}$}
\end{figure}

The resonant circulating intensity, $\text{I}$, in a high-$\text{Q}$ and small-$\text{V}$ optical resonator can be estimated~\cite{Vahala} as $\text{I} = \text{P}_\text{in} \left( \frac{\lambda}{2 \pi \text{n}_\text{g}}  \right) (\text{Q}/\text{V})$, where $\text{P}_\text{in}$ is the input excitation power at wavelength $\lambda$ and $\text{n}_\text{g}$ is the group index of the resonant mode.\ For an input power of $1\,\text{mW}$ and a typical Anderson-localized mode with $\text{Q} \sim 10^5$ and $\text{V}=1.2(\lambda/n)^3$, this corresponds to $7.2\,\text{GW}/\text{m}^2$.\ This large stored energy results in a significant non-linear behaviour of the dielectric medium.\ Due to the central symmetry in silicon, the susceptibility tensor of second order is null and only third order non-linear terms need to be considered.\ To such order, the main non-linear processes in silicon for single frequency operation and in the telecom frequency range are two-photon absorption and a dispersive Kerr effect, arising from the real and imaginary parts of the third order susceptibility.\ As a large free carrier population $\text{N}_\text{e}$ can be generated in such high-Q/low-V cavities, the further absorption of free carriers needs also to be considered.\ Most of the absorbed optical power in the cavity is released to the lattice through the decay of the photoexcited carriers, rising its temperature.\ As this decay is much faster than the temperature dynamics, the energy transfer from the electron population to the lattice can be considered instantaneous when modelling the temperature field in the cavity region.\ Importantly, the dispersion on the cavity resonant frequency $\omega_{\text{c}}$ induced by this temperature rise is opposed to the one mediated by the presence of free carriers, which is the key to enter a periodic temporal modulation of the optical cavity under certain conditions.\ Following the derivations in Refs.~\cite{nonlinear1,nonlinear2,nonlinear3}, all these non-linear processes can be introduced microscopically into the Maxwell's equations with the corresponding non-linear polarization terms, expressed in terms of coupled-mode formalism as:

\begin{equation}
 \label{CPTnonlinear1}
\dot{A}= -\left(i\Delta + \frac{\kappa}{2} + \frac{c^2}{n_{\text{Si}}^2}\frac{\beta_\text{TPA}\lvert A\rvert^2}{2\text{V}_\text{TPA}}+ \frac{c}{n_{\text{Si}}}\frac{\sigma_{r}\text{N}_\text{e}}{2 \text{V}_\text{FCA}} \right) A -\sqrt{\frac{\kappa_\text{e}}{2}\text{P}_{\text{in}}}
\end{equation}
\begin{equation}
\label{CPTnonlinear2}
\Delta= \omega_{\text{c}} + \frac{\omega_{\text{c}}}{n_{\text{Si}}}\frac{\sigma_{i}\text{N}_\text{e}}{\text{V}_\text{FCA}} - \frac{\omega_{\text{c}}}{n_{\text{Si}}}n_{\text{T}}\Delta \text{T} - \omega_{\text{L}}
\end{equation}
where the two-photon absorption and free carrier absorption are explicitly included in In (\ref{CPTnonlinear1}), while the dispersion due to the excess of free carriers $\text{N}_\text{e}$ and the heating of the structure $\Delta\text{T}$ are also pointed out in (\ref{CPTnonlinear2}).\ The different parameters included in these equations are: $\beta_\text{TPA}$ the tabulated two-photon absorption coefficient, $\text{V}_\text{TPA}$ and $\text{V}_\text{FCA}$ the characteristic two-photon and free-carrier absorption volumes, respectively, $n_{\text{Si}}$ the silicon refractive index, c the speed of light, $\sigma_r$ and $\sigma_i$ the free-carrier absorption and dispersion \textit{cross sections} and $n_\text{T}$ the dependence of the index of refraction on the temperature of the structure.\ Here, the dispersion associated to the Kerr effect has been neglected since it is negligible compared to the dispersion induced by the free carriers and the structural heating.\\

The dynamics of the system observed in the manuscript are fully determined by the evolution of the free carrier population, $\text{N}_\text{e}$, and the temperature variation, $\Delta\text{T}$.\ Although the evolution of these quantities is complex, it is possible to give some physical insight including the absorption processes mentioned above and some phenomenological decay rates as:
\begin{equation}\label{FCD}
\frac{\mathrm{d}\text{N}_\text{e}}{\mathrm{d}t} = -\gamma_\text{fc} \text{N}_\text{e} + \frac{1}{2}\frac{c^2}{n_\text{Si}^2}\frac{\beta_\text{TPA}\lvert A\rvert^4}{\hbar\omega_\text{L}\text{V}_\text{TPA}}+\frac{c}{n_\text{Si}}\frac{\alpha_\text{lin}\lvert A\rvert^2}{\hbar\omega_{\text{L}}R_\text{eff}}
\end{equation}
\begin{equation}\label{TO}
\frac{\mathrm{d}\Delta \text{T}}{\mathrm{d}t} = -\gamma_\text{th} \Delta \text{T} + \frac{1}{\rho_\text{Si}C_\text{p,Si}\text{V}_\text{eff,T}} \left(\frac{c^2}{n_{Si}^2}\frac{\beta_\text{TPA}\lvert A\rvert^4}{\text{V}_\text{TPA}}+\frac{c}{n_\text{Si}}\frac{\sigma_{r}\lvert A\rvert^2 \text{N}_\text{e}}{\text{V}_\text{FCA}}+\frac{c}{n_\text{Si}}\frac{\alpha_\text{lin}\lvert A\rvert^2}{R_\text{eff}}\right)
\end{equation}
here, a linear absorption, $\alpha_\text{lin}$, is considered in addition to the two-photon and free carrier absorption processes.\ The effect of this linear term is included implicitly in (\ref{CPTnonlinear1}) through the overall photonic decay rate $\kappa$.\ Here, the different parameters are $R_\text{eff}$, which is the inverse of the fraction of the optical mode inside the structure, $\rho_\text{Si}$ and $C_\text{p,Si}$ the silicon density and silicon specific heat capacity at constant pressure, $\text{V}_\text{eff,T}$ a cavity effective thermal volume and $\gamma_\text{fc}$ and $\gamma_\text{th}$ the free-carrier and thermal decay rates.

The decay time of most of the Anderson-localized modes observed in this work ($\sim \text{ps}$) is orders of magnitude faster than the nonlinear dispersion mechanisms listed here ($\sim \text{ns}$).\ Therefore, $\text{N}_\text{e}$ and $\Delta \text{T}$ can be considered constant within the time-scales involved in (\ref{CPTnonlinear1}).\ Additionally, the losses induced by the non-linear processes can be assumed to be negligible with respect to the linear ones.\ Considering this adiabatic response of the optical cavity with the steady state number of photons (\ref{no}), the detuning given by (\ref{CPTnonlinear2}) and $\rvert A\lvert^2=\hbar\omega_{\text{L}}n_\text{ph}$, the equations (\ref{FCD}) and (\ref{TO}) result in the simplified system of non-linear equations:

\begin{equation}\label{FCDdef}
\frac{\mathrm{d}\text{N}_\text{e}}{\mathrm{d}t} = -\gamma_\text{fc} \text{N}_\text{e} + \frac{1}{2}\frac{c^2}{n_\text{Si}^2}\frac{\beta_\text{TPA}\hbar\omega_{L}n_{ph}^2}{\text{V}_\text{TPA}}+\frac{c}{n_\text{Si}}\frac{\alpha_\text{lin} n_\text{ph}}{R_\text{eff}}
\end{equation}
\begin{equation}\label{TOdef}
\frac{\mathrm{d}\Delta \text{T}}{\mathrm{d}t} = -\gamma_\text{th} \Delta \text{T} + \frac{\hbar\omega_\text{L}n_\text{ph}}{\rho_\text{Si}C_\text{p,Si}\text{V}_\text{eff,T}} \left(\frac{c^2}{n_\text{Si}^2}\frac{\beta_\text{TPA}\hbar\omega_\text{L} n_\text{ph}}{\text{V}_\text{TPA}}+\frac{c}{n_\text{Si}}\frac{\sigma_{r}\text{N}_\text{e}}{\text{V}_\text{FCA}}+\frac{c}{n_\text{Si}}\frac{\alpha_\text{lin}}{R_\text{eff}}\right)
\end{equation} \\
where $n_\text{ph}$ is given by (\ref{no}) with the detuning in (\ref{CPTnonlinear2}).\ These equations are solvable with less computational effort while being still able to capture most of the experimental features in our experiments and previous experiments in engineered cavities~\cite{dani_selfpulsing}.\ For modes with very large $\text{Q}$-factor, the assumption of adiabaticity does not hold which requires to solve the system of equations (\ref{CPTnonlinear1}), (\ref{CPTnonlinear2}), (\ref{FCD}) and (\ref{TO}) instead.\\

\begin{figure}[t!]
  \includegraphics[width=15cm]{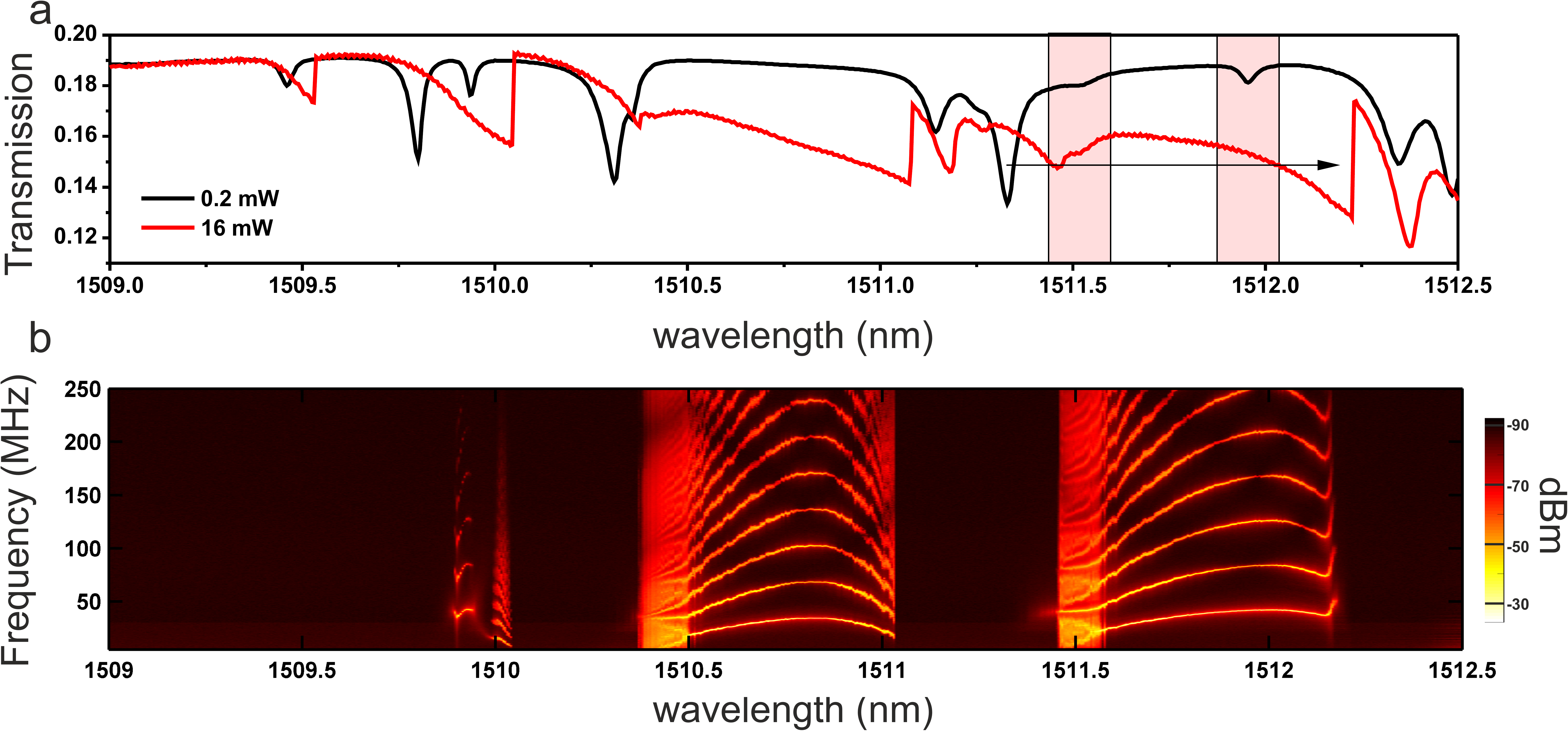}
    \caption{ \label{S7} (Color online) (a) Transmission trough a tapered fiber measured while scanning the input laser wavelength at an input power $\text{P}_\text{in}=0.2\,\text{mW}$ (black line) and $\text{P}_\text{in}=16\,\text{mW}$ (red line).\ (b) Color contour plot of the radiofrequency power spectral density at an input power of $16\,\text{mW}$ while scanning the input laser wavelength through three different Anderson-localized mode with a \textit{cold} wavelengths at $\lambda_\text{c}=1509.79\,\text{nm}$, $1510.3\,\text{nm}$ and $1511.32\,\text{nm}$ (plotted in a).}
\end{figure}

In most of the situations, the dynamic solution to the system of differential equations (\ref{FCDdef}) and (\ref{TOdef}) is a fixed or equilibrium point in the phase space $\{\text{N}_\text{e},\Delta \text{T}\}$ which leads to a fixed and stable spectral shift of the cavity mode.\ For particular combinations of $\text{P}_\text{in}$ and $\lambda_\text{L}$, the dynamic solution forms a stable closed trajectory in the phase space $\{\text{N}_\text{e},\Delta \text{T}\}$ or limit cycle~\cite{nonlinear1} which is generally highly anharmonic.\ This phenomenon has been observed in different photonic engineered nanostructures such as microdisks~\cite{nonlinear1}, photonic crystals~\cite{nonlinear_phot_crystal} or optomechanical crystals~\cite{dani_selfpulsing}.\ In our system, this happens when the number of cavity photons $\text{n}_\text{ph}$ is above a threshold value $n_\text{ph,th}$.\ Integrating numerically Eqs.~(\ref{FCDdef})-(\ref{TOdef}) with the parameters summarized in the table~\ref{table} reproduces accurately the temporal modulation measured in our experiments and the radiofrequency spectra obtained while sweeping the laser wavelength, as shown in Fig.\ref{S6}.\ Despite the rather good agreement obtained between our simplified model, Fig.\ref{S6}(c), and the experiment, Fig.\ref{S6}(d), there are some features that cannot be captured by the model.\ In particular, the blue-detuned flank of the spectral scanning of the mode, which corresponds to the bright region in Fig.~\ref{S6}(d) spanning $1510.35$ - $1510.5\,\text{nm}$.

\begin{table}[ht]
\caption{Model parameters}
\centering
\begin{tabular}{|c|c|c|c|}
  \hline
  Parameter & Value & Units & Origin \\
  \hline
Refractive index ($n_\text{Si}$) & 3.458 & - & ~\cite{silicon} \\
Density ($\rho_\text{Si}$) & $2.33\cdot 10^{3}$ & $kg/m^3$ & ~\cite{silicon} \\
Constant-pressure specific heat capacity ($C_\text{p,Si}$) & $0.7\cdot 10^{3}$ & $J/(kg\cdot K)$ & ~\cite{silicon} \\
Two-photon absorption coefficient ($\beta_\text{TPA}$) & $8.4\cdot 10^{-12}$ & m/W & ~\cite{betaTPA}  \\
Free-carrier absorption cross-section ($\sigma_{r}$) & $1.45\cdot 10^{-21}$ & $m^2$ & ~\cite{sigma} \\
Free-carrier dispersion cross-section ($\sigma_{i}$) & -$5.3\cdot 10^{-27}$ & $m^2$ & ~\cite{sigma} \\
Linear absorption coefficient ($\alpha_\text{lin}$) & 3.68 & $m^{-1}$ & ~\cite{alphalin} \\
Refractive index temperature coefficient ($n_\text{T}$) & $1.86\cdot 10^{-4}$ & $\text{K}^{-1}$ & ~\cite{nT} \\
Thermal decay rate ($\gamma_\text{th}$) & 33 & MHz & fit \\
 Free-carrier decay rate ($\gamma_\text{fc}$) & 4.62 & GHz & fit\\
Effective thermal volume ($V_\text{eff,T}$) & $2.3\cdot 10^{-18}$ & $ m^3$ & fit \\
Free-carrier absorption volume ($V_\text{FCA}$) & $8\cdot 10^{-18}$ & $m^3$ & fit \\
Two-photon absorption volume ($V_\text{TPA}$) & 7 & $(\lambda_{c} / n_{Si})^3$ & ~\cite{Vasco}\\
  \hline
\end{tabular}
\label{table}
\end{table}

A final comment on the interaction between structural complexity, interference and material nonlinearities is required.\ The interplay between nonlinear interactions and wave interference in strongly scattering media is complex and subtle.\ Figure~\ref{S6}(a) plots the transmission spectrum trough a tapered fiber at a field position along the waveguide and at two different input laser powers well bellow ($\text{P}_\text{in}=0.2\,\text{mW}$, black line) and well above ($\text{P}_\text{in}=16\,\text{mW}$, red line) the threshold cavity photon number $n_\text{0,th}$.\ The spectral shift induced on different modes when increasing the laser input power is significantly mode-dependent which allows tuning different modes on resonance for particular values of the input laser wavelength.\ As shown in Figure~\ref{S7}(a), the mode at $\lambda_\text{c}=1511.32\,\text{nm}$ for $\text{P}_\text{in}=0.2\,\text{mW}$ is spectrally tuned through two different modes with cold wavelengths $\lambda_\text{c}=1511.51\,\text{nm}$ and $\lambda_\text{c}=1511.96\,\text{nm}$, respectively.\ However, this spatial and spectral mode overlapping results in temporal traces that can be explained by a single-mode model as described by Eqs.~\ref{FCD} and~\ref{TO}.\ In particular, Figure~\ref{S7}(b) plots the contour plot of the radiofrequency power spectral density for an input power of $\text{P}_\text{in}=16\,\text{mW}$.\ It is possible to identify three Anderson-localized modes entering a limit cycle which correspond to the modes with a \textit{cold} wavelength of $\lambda_\text{c}=1509.79\,\text{nm}$, $\lambda_\text{c}=1510.3\,\text{nm}$ and $\lambda_\text{c}=1511.32\,\text{nm}$.\ Although other modes are on resonance with these ones for different input laser wavelengths, neither the radiofrequency contour plot nor their temporal traces shows any effect of mode interaction.\ Mode interaction is predicted to be limited by localization in nonlinear complex systems, as studied in multimode lasers for modes with significant spatial or frequency overlap~\cite{interactions}.

      \end{widetext}

\end{document}